\documentclass[conference,10pt,a4paper]{IEEEtran}

\usepackage[left=1.62cm,right=1.62cm,top=1.9cm]{geometry}
\usepackage[acronyms]{glossaries}
\newacronym{icas}{ISAC}{integrated sensing and communications}
\newacronym[plural=UAVs]{uav}{UAV}{unmanned aerial vehicle}
\newacronym{hrpe}{HRPE}{high-resolution parameter estimation}
\newacronym[plural=GPSDOs]{gpsdo}{GPSDO}{global positioning system disciplined oscillator}
\newacronym{los}{LoS}{line-of-sight}
\newacronym[plural=GNSSs]{gnss}{GNSS}{global navigation satellite system}
\newacronym[plural=CFOs]{cfo}{CFO}{carrier frequency offset}
\newacronym[plural=SFOs]{sfo}{SFO}{sampling frequency offset}
\newacronym{sto}{STO}{symbol time offset}
\newacronym{rtk}{RTK}{real time kinematic}
\newacronym{ofdm}{OFDM}{orthogonal frequency-division multiplexing}
\newacronym{lo}{LO}{local oscillator}
\newacronym[plural=RMSEs]{rmse}{RMSE}{root mean square error}
\newacronym{cdf}{CDF}{cumulative distribution function}
\newacronym{mimo}{MIMO}{multiple-input and multiple-output}
\usepackage{siunitx}
\usepackage{amsfonts}
\usepackage{cleveref}
\usepackage[figurename=Figure]{caption}
\usepackage{subcaption}
\usepackage{booktabs}
\usepackage[activate={true,nocompatibility},final,tracking=true,kerning=true,spacing=true,factor=1100,stretch=30,shrink=30]{microtype} % seb used 30 30
\usepackage{tikz}
\usepackage{pgfplots}
\usetikzlibrary{calc}
\usetikzlibrary{plotmarks}
\DeclareUnicodeCharacter{2212}{−}
\pgfplotsset{compat=newest}
\usepackage{amsmath}
\usepackage{amssymb}

\DeclareMathOperator*{\argmin}{argmin}
\DeclareMathOperator*{\argmax}{argmax}

\title{Geometry-Based Drift Compensation\\ for Distributed Channel Sounding Measurements\\ in Dynamic Drone Scenarios}
\author{
    \IEEEauthorblockN{
        Lorenz~Mohr\IEEEauthorrefmark{1},
        Marc~Miranda\IEEEauthorrefmark{1},
        Sebastian~Semper\IEEEauthorrefmark{1}\IEEEauthorrefmark{2},
        Julia~Beuster\IEEEauthorrefmark{1},\\
        Carsten~Andrich\IEEEauthorrefmark{1},
        Sebastian~Giehl\IEEEauthorrefmark{1},
        Christian~Schneider\IEEEauthorrefmark{1}\IEEEauthorrefmark{2},
        and
        Reiner~S.~Thom\"a\IEEEauthorrefmark{1}
    }
    \IEEEauthorblockA{
        \IEEEauthorrefmark{1} Technische Universit\"at Ilmenau, Insitute for Information Technology, Ilmenau, Germany
    }
    \IEEEauthorrefmark{2}
    Fraunhofer Institute for Integrated Circuits IIS, Ilmenau, Germany
}

\usepackage[style=ieee,backend=biber,sorting=none, url=false, doi=true]{biblatex}
\DeclareFieldFormat[thesis]{url}{[Online]. \mkbibacro{Available:}\space\url{#1}}

\AtEveryBibitem{
    \clearlist{address}
    \clearfield{month}
    \clearfield{editor}
    \clearfield{eprint}
    \clearfield{volume}
    \clearfield{isbn}
    \clearfield{issn}
    \clearlist{location}
    \clearlist{editor}
    \clearfield{pages}
}
\renewcommand{\baselinestretch}{0.95}
\begin{document}
\maketitle
% reduce space around equations
%\abovedisplayskip      = 1pt plus 1pt minus 1pt
%\abovedisplayshortskip = 1pt plus 1pt minus 1pt
%\belowdisplayskip      = 1pt plus 1pt minus 1pt
%\belowdisplayshortskip = 1pt plus 1pt minus 1pt
% input sections
\begin{abstract}
    \sisetup{propagate-math-font = true, reset-math-version = false}
    %%% OFFICIAL VERSION
    Measured impulse responses obtained from a dynamic \gls{uav} channel sounding system exhibit effects attributable to time-varying \gls{cfo} and \gls{sfo}.
    %We observed time-varying \glspl{cfo} and \glspl{sfo} within mobile multi-node channel sounding measurements.
    %On the other hand, performing reliable delay-Doppler parameter estimation from channel sounding data generally requires coherency among the nodes, implying a continuously differentiable phase progression of the received signals.
    To correct the recorded data in post-processing, we extend existing geometry-based drift compensation algorithms by an explicit \gls{los} determination---combining a symbol-wise \gls{hrpe} in delay with a Kalman filter.
    This proposed extension facilitates the removal of rapidly varying synchronization mismatches from channel sounding measurements in rich multipath propagation scenarios.
    %This processing step simultaneously improves the robustness of the \gls{los} estimate in multipath propagation and enables the correction of the rapidly varying synchronization mismatches while preserving relative sensor motion.
    Furthermore, we propose using the relative residual power after subtraction of estimated multipath components as a metric for ground-truth-independent comparison of post-processing synchronization methods for recorded channel sounding data.
    The application of the proposed procedure shows that our approach outperforms existing post-processing compensation algorithms, reducing the relative residual power by more than~\boldmath{\qty{5}{\decibel}} and the delay-Doppler estimate \glspl{rmse} of a passive \gls{uav} target by approximately~\boldmath{\qty{60}{\percent}}.
\end{abstract}
\glsresetall
\section{Introduction}
\label{1-intro}

% old version
%Communication networks utilize a multitude of distributed nodes to cover a physically large area with minimal blind spots.
%If these networks provide \gls{icas} functionalities, some of the nodes can be mounted on a \glspl{uav} to tackle search and rescue or surveillance and security.
%The resulting \gls{uav} channels feature unique properties and, therefore, necessitate synchronized multisensor channel measurements for future channel model standardization and algorithmic development~\cite{2024_survey_uav_channel_sounding_Mao}.
%However, the wireless synchronization of these sensors is non-trivial and requires special procedures.
%Following the survey in~\cite{2024_survey_uav_channel_sounding_Mao}, \glspl{gpsdo} constitute one option to perform this synchronization task.
%As these devices derive their reference signals from highly accurate \glspl{gnss}, they provide short- and long-term stable time and frequency references, achieving \unit{\nano\second}-level synchronization across the sounding nodes.

% REVISION LOMO
% short general motivation of the paper
Communication networks utilize a multitude of distributed nodes to cover a physically large area with minimal blind spots.
If these networks also provide sensing functionalities, some of the nodes can be mounted on \glspl{uav} to tackle search and rescue or surveillance and security.
The resulting \gls{uav} channels feature unique properties and, therefore, necessitate multi-node channel measurements to extract these characteristics~\cite{2024_survey_uav_channel_sounding_Mao}.

Performing accurate frequency and time synchronization is crucial for these distributed channel sounding measurement systems. %, especially if some of the nodes are \gls{uav}-mounted.
In particular, all transmitters and receivers of such a system require common time and frequency references.
Following the survey in~\cite{2024_survey_uav_channel_sounding_Mao}, \glspl{gpsdo} constitute one option to accomplish this objective.
As these devices derive their reference signals from highly accurate \glspl{gnss}, they provide short- and long-term stable time and frequency references, theoretically achieving \unit{\nano\second}-level synchronization across the sounding nodes.

% our observation of the data
However, we found limitations in the synchronization accuracy of \glspl{gpsdo}.
In essence, hardware-constrained time pulse accuracy and external effects like temperature fluctuation, mechanical disturbance, or \gls{gnss} signal impairment impose drifts on the reference signals of the \glspl{gpsdo}~\cite{2024_gpsdo-characterization-uav_Beuster}.
As each node derives its local oscillator and sampling frequencies from these signals, any drift in the references manifests as \gls{cfo}, \gls{sfo}, and \gls{sto} in the channel data.
We observe these effects in the datasets obtained when using the multi-node channel sounding system presented in~\cite{2025_testbed_icas4m_measurement_Beuster}.
In essence, \gls{cfo}, \gls{sfo}, and \gls{sto} result in non-smooth phase progression and a time-variant drift of the estimated \gls{los} from its \gls{gnss}-\gls{rtk}-derived position ground truth.

% introducing the term coherency / incoherency and motivation for correction
As noted in~\cite{wang_system_error_2022}, compensating for such system calibration errors enhances the usability of channel sounding datasets in applications that rely on coherent phase information---delay-Doppler estimation, Fourier-transform-based processing, or background subtraction---for example.
According to~\cite{2025_distributed_multisensor_isac_Thomae}, the term coherent implies a smoothly varying and continuously differentiable phase progression across the received signal.
Due to \gls{gpsdo}-internal phase tracking and correction algorithms, the phase progression of the measurements, however, demonstrates to be not smooth and shows several corners yielding incoherencies.
Hence, a compensation of time-varying \gls{cfo}, \gls{sfo}, and \gls{sto} promotes the utilization of the multi-static sounding data for applications like channel characterization or passive target localization.

% sota review
% transmit signal structure
There already exist several approaches in the literature dealing with post-processing synchronization in \gls{ofdm}-based multi-node systems.
Some algorithms utilize special transmit signal structures that enable the estimation and correction of \gls{cfo}, \gls{sfo}, and \gls{sto} at the receiver.
A classical example is the Moose algorithm which estimates time-variant \gls{cfo} by repeating the transmitted symbol~\cite{1994_cfo-correction_Moose}.
While computationally efficient, this method is limited to \gls{cfo} estimation and does not account for \gls{sfo} and \gls{sto}.
The Schmidl~\&~Cox algorithm extends this concept by introducing a dedicated \gls{ofdm} preamble structure that allows for joint estimation of \gls{cfo} and \gls{sto}~\cite{1997_ofdm-freq-time-sync_Schmidl}.
More recent work, such as~\cite{2023_ofdm-synchronization_deOliveira}, integrates an additional \gls{sfo} estimation stage into the Schmidl~\&~Cox algorithm to mitigate residual synchronization mismatches.

% why we need geometry-based
While these approaches prove to be powerful tools for drift compensation, they require a pre-designed measurement signal structure and are therefore not suitable for transmit-signal-agnostic post-processing of multi-sine-based channel sounding data.
In addition, these algorithms estimate the synchronization offsets on a coherent processing interval level, thereby inherently assuming these effects to drift on a timescale much longer than this interval duration.
This assumption is, hence, not valid for our and comparable measurement systems~\cite{2025_testbed_icas4m_measurement_Beuster, wang_system_error_2022}.

Targeting methods that are transmit-signal-agnostic,~\cite{wang_system_error_2022} describes a method for error calibration of published datasets affected by time-varying \gls{sto} and \gls{cfo}.
The procedure employs the first recorded symbol as a reference to correct timing and phase drifts in subsequent ones.
This compensation, however, discards information about the relative motion between transmitter and receiver.
In addition, the authors apply grid searches to estimate \gls{cfo} and \gls{sto}, which further constrain the correction accuracy to the chosen grid granularity.
To overcome this limitation, a grid-free approach should be employed instead.

In~\cite{2022_geometry_based_mimo_synchronization_Euchner}, Euchner introduces a geometry-based algorithm for \gls{sfo} and \gls{cfo} compensation for dynamic \gls{mimo} channel sounding data.
By comparing the measured channel frequency responses to the expected \gls{los} response, which is derived from accurate transmitter and receiver positions, he estimates \gls{sfo} and \gls{cfo} correction values.
These corrections compensate the sounding data such that it is similar to the \gls{los} response thus maintaining the relative movement between transmitter and receiver.
However, this approach has two major shortcomings rendering its application limited for post-processing of our measurements.
First, the comparison of the complete channel frequency response to that of the \gls{los} inherently assumes that this path dominates the response.
Second, Euchner assumes time-invariant \gls{sfo} and \gls{cfo}, enabling the aggregation of several symbols for offset estimation.
These assumptions, however, are not satisfied by the planned measurement scenarios with our sounding system~\cite{2025_testbed_icas4m_measurement_Beuster}.
Due to the dynamic outdoor scenario including fast moving nodes, the channel features rich multipath propagation with multiple dominant paths and \gls{los} fading.
As a result, the \gls{los} path does not always dominate the measurements, yielding incorrect compensation values.
Moreover, the corresponding offsets in the sounding data are time-variant.
These two circumstances prohibit the aggregation of consecutive channel sounding measurements for correction and instead require an explicit and symbol-wise \gls{los} estimation for \gls{cfo} and \gls{sfo} compensation.
%Burmeister performs an interpolation of the channel impulse response to remove the influence of \gls{sfo} from the amplitude of the \gls{los}, neglecting phase effects, which are, however, crucial for an accurate estimation of this path in multipath propagation~\cite{2021_fractional-sampling-time-offsets_Burmeister}. 

%%% MERGE THESE TWO THINGS INTO OUR 
Our contributions to geometry-based drift compensation of channel sounding measurements are twofold.
To combat multipath propagation effects, we explicitly estimate the \gls{los} on a symbol level using a \gls{hrpe} algorithm followed by a Kalman filter.
Overall, this combination enables the geometry-based correction of time-varying drifts by obtaining a \gls{los} estimate that is robust to multipath fading and spurious detections of the underlying \gls{hrpe} algorithm.
%Based on these estimates, we determine the \gls{los}, which is not as trivial as selecting either the multipath component with the lowest propagation delay or the highest power.
%Due to spurious detections, strong ground reflections, temporary \gls{los} shadowing, and limited knowledge of the antenna radiation pattern, we observe that both of these heuristics fail to reliably determine the propagation parameters---delay and phase---of the \gls{los}.
%Instead, we employ a Kalman filter for reliable \gls{los} estimation based on the multipath delays.
On the other hand, we propose the use of a metric to determine the quality of the correction achieved by different compensation procedures.
This metric, the relative residual power, determines how well a model-based \gls{hrpe} algorithm describes the data by an analytical signal model.
Since these algorithms provide channel parameters by assuming perfect synchronization, any synchronization mismatches result in wrong parameters and thus an erroneous description of the measurements.
Consequently, the relative residual power allows to quantify synchronization errors.
\section{Preliminaries}
\label{2-basics}

% describe the setup briefly and introduce ofdm system parameter table
\subsection{Channel Sounding System}
\label{2-basics/channel-sounding-setup}

The uncompensated dataset at hand stems from a multi-static distributed measurement system comprising one \gls{uav}-based transmitting and seven receiving nodes, three \gls{uav}-based and four stationary~\cite{2025_testbed_icas4m_measurement_Beuster}.
During the measurement, the \gls{uav}-based nodes exhibited a dynamic behavior.
%\Cref{fig:2-meas-setup} depicts the spatial configuration and trajectories of the eight active nodes and one passive target \gls{uav}.
All active and passive nodes were equipped with a \gls{gnss}-\gls{rtk} unit recording their location during channel sounding with \si{\centi\meter}-level precision~\cite{2025_testbed_icas4m_measurement_Beuster}.
The available position information constitute the geometric ground truth and enable the subsequent performance evaluation of node synchronization and passive target localization.

To achieve distributed synchronization across nodes, the channel sounding system utilizes \glspl{gpsdo}.
Under laboratory conditions, these devices establish one common time and frequency reference at each node thus yielding synchronized measurements between the transmitter and the receivers~\cite{2025_testbed_icas4m_measurement_Beuster}.

%In fact, the latter requires channel state information of multiple highly synchronized nodes~\cite{2025_distributed_multisensor_isac_Thomae}.

%
%\begin{figure}
%    \centering
%    \missingfigure{meas setup configuration}
%    \caption{Caption}
%    \label{fig:2-meas-setup}
%\end{figure}
%

\subsection{Signal Model}
\label{2-basics/signal-model}

To estimate the channel frequency response, the transmitter illuminates the environment by sending $L \in \mathbb{N}$ symbols comprising $K \in \mathbb{N}$ subcarriers.
\Cref{tab:2-signal-parameters} lists the corresponding parameters of the transmitted sounding signal.
The receiver acquires these signals through a number of specular propagation paths, each encompassing a distinctive propagation delay, Doppler-shift, and complex path weight.
Assuming the \glspl{gpsdo} achieve perfect synchronization between nodes, the received signal is filtered, down-converted to complex baseband, and sampled in time and frequency with the corresponding sampling intervals~$1/f_\mathrm{s}$ and~$\Delta f$.

\begin{table}[h]
\sisetup{propagate-math-font = true, reset-math-version = false}
\centering
\caption{\textit{Parameters of the OFDM Signal}}
\begin{tabular}{c c c}
Signal Parameter & Symbol & Value\\
\toprule
Carrier Frequency & $f_\mathrm{c}$ & \qty{3.75}{\giga\hertz}\\
Symbol Length & $\Delta t$& \qty{16}{\micro\second}\\
Bandwidth (used) & $f_\mathrm{s}$ & \qty{80}{\mega\hertz} (\qty{48}{\mega\hertz})\\
Carrier Spacing & $\Delta f$ & \qty{62.5}{\kilo\hertz}\\
Subcarriers (used) & $K$ & \num{1280} (\num{768})\\
\bottomrule
\end{tabular}
\label{tab:2-signal-parameters}
\end{table}

Performing a point-wise division of the received samples by the transmitted signal and back-to-back calibration data in the frequency domain yields the complex baseband channel frequency response $\mathbf{H} \in \mathbb{C}^{K \times L}$ between one transmitter-receiver link, where
\begin{equation}
    \label{eq:2-signal-model}
    \mathbf{H}_{k\ell}=
    \sum_{p=1}^{P} \gamma_p \cdot \mathrm{e}^{\mathrm{j} 2\pi \alpha_p \ell} \cdot \mathrm{e}^{-\mathrm{j} 2\pi \tau_p k} + \mathbf{N}_{k\ell},
\end{equation}
denotes the element at the~\mbox{$k$-th} subcarrier and $\ell$-th symbol.
Here, $P  \in \mathbb{N}$ is the total number of propagation paths and $\gamma_p \in \mathbb{C}$, $\alpha_p \in \mathbb{R}$, and $\tau_p \in \mathbb{R}$ are the complex weight, Doppler-shift, and propagation delay of the $p$-th path, respectively.
The complex path weight represents path loss, path phase, and angle- and polarization-dependent antenna beam pattern information
In addition, Doppler-shift and propagation delay are normalized by their maximum unambiguous values $1/t_\mathrm{s}$ and $1/\Delta f$.
Furthermore, the elements of $\mathbf{N} \in \mathbb{C}^{K \times L}$ are assumed to be drawn from an independent and identically distributed zero-mean circularly symmetric Gaussian process and account for noisy observations of the channel.
%In essence, propagation delay and Doppler-shift of each specular path result in a linear growing phase across subcarriers and \gls{ofdm} symbols.

\subsection{Observed Effects in the Measurements}
\label{2-basics/observed-effects}

Despite several measures to achieve node synchronization of the channel sounding system~\cite{2025_testbed_icas4m_measurement_Beuster}, we noticed synchronization mismatches in the measured frequency responses, which can be attributed to time-variant drifts in the \glspl{gpsdo} reference signals~\cite{2024_gpsdo-characterization-uav_Beuster}.
%In addition, the \gls{gpsdo} repeatedly correct their reference signals using internal control loops. CITE REFERENCE GPSDO CORRECTION
As each node derives its \gls{lo} and sampling clock frequencies from these devices, drifts in the references introduce time-varying \gls{cfo} and \gls{sfo} to the measurements.
Consequently, the sounding data comprise not only the propagation channel modeled by~\eqref{eq:2-signal-model} but also the dynamic effects of the measurement system that cannot be removed by offline back-to-back calibration.

\begin{figure}[hb]
    \centering
    \begin{subfigure}{1\columnwidth}
    \centering
    \includegraphics[width=0.9\columnwidth]{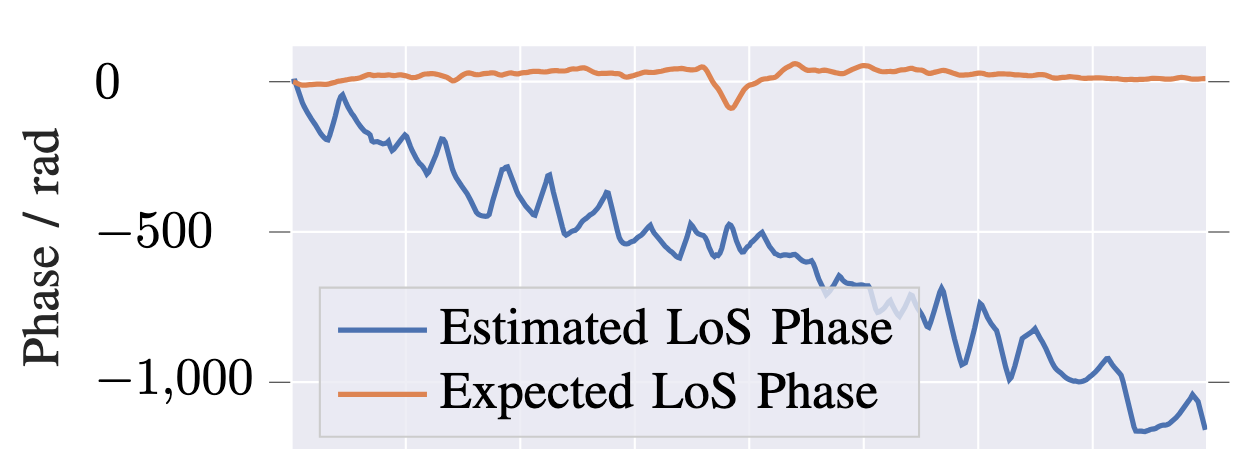}
    \vspace{0em}
    \caption{\textit{Phase Offset from CFO}}
    \label{fig:2-offset-phase}
    \end{subfigure}
    %\hfill
    \begin{subfigure}{1\columnwidth}
    \centering
    \includegraphics[width=0.9\columnwidth]{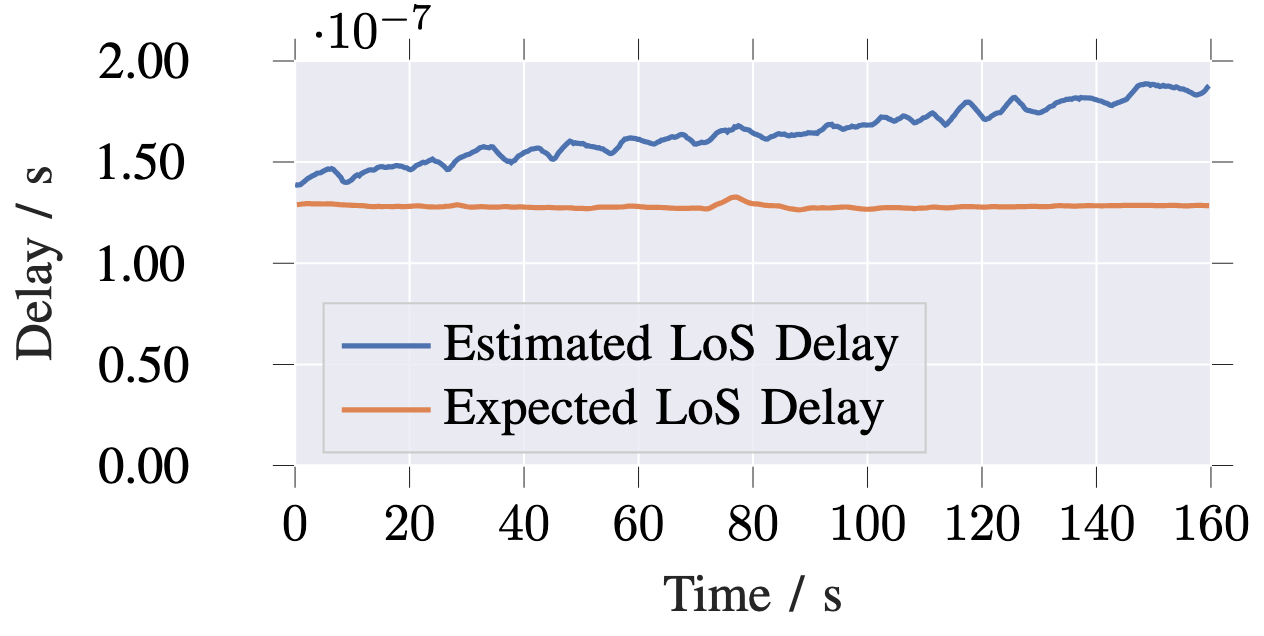}
    \caption{\textit{Delay Offset from SFO}}
    \label{fig:2-offset-delay}
    \end{subfigure}
    \caption{\textit{Drift of the Unwrapped LoS Phase and Delay---The uncompensated data shows a drifting and non-smooth phase progression~(a) and an increasing offset in the LoS delay~(b). We derive both expected values geometrically utilizing the known transmitter and receiver positions.}}
    \label{fig:2-offset}
\end{figure}

By naively applying an \gls{hrpe} algorithm to the uncompensated data, the observable effects of \gls{cfo} and \gls{sfo} in this superposition are twofold.
First, \gls{cfo} alters the phase progression of the \gls{los} component.
%Since the measurement setup was designed such that the \gls{los} is a dominant path for most of the time, we would have expected a phase progression for the sounding data that matches the phase of the \gls{los}.
The unwrapped phase of this path exhibits a progressions that differ from the geometrically-derived expected \gls{los} phase, as \Cref{fig:2-offset-phase} illustrates.
In particular, the estimated phase progression is not continuously differentiable and the corners yield incoherencies during delay-Doppler processing (\Cref{fig:4-ddmap-smear}).
On the other hand, \gls{cfo} and \gls{sfo} induce drifts in the delay and Doppler parameter space, respectively.
As \Cref{fig:2-offset-delay} depicts, the offset between the expected and the estimated  \gls{los} grows over time.
Consequently, the results of \gls{hrpe} delay-Doppler processing do not align with the ground truth (\Cref{fig:4-ddmap-shift}).

%To render the recorded channel sounding applicable for further processing and evaluation, it is therefore crucial to estimate and remove the observed synchronization mismatches.
%Since the subsequent processing steps are receiver-agnostic, we drop the receiver index $i$ for conciseness.

%

\subsection{Synchronization Error Modeling}
\label{2-basics/synchronization-error-modeling}

As discussed in \Cref{2-basics/observed-effects}, \gls{cfo} and \gls{sfo} distort the ideal frequency response~\eqref{eq:2-signal-model}.
A time-varying \gls{cfo} relates to the fact that the carrier frequencies at transmitter and receiver differ.
According to~\cite{1994_cfo-correction_Moose}, this difference manifests as an additional phase shift over time, denoted by the complex phasor
\begin{equation}
    \label{eq:2-cfo}
    \Psi_\text{CFO}[\ell] = \mathrm{e}^{\mathrm{j} 2\pi \ell \mu[\ell]}.
\end{equation}
Here, $\Psi_\text{CFO}[\ell] \in \mathbb{C}$ and the quantity $\mu[\ell] \in \mathbb{R}$ represents the time-varying carrier frequency difference between transmitter and receiver normalized by $\Delta f$.

Likewise, a time-varying \gls{sfo} yields a mismatch of the sampling frequencies at transmitter and receiver.
As we assume \gls{sfo} to be sufficiently small, it only introduces a phase shift across subcarriers, termed \gls{sto}~\cite{2022_geometry_based_mimo_synchronization_Euchner}.
The corresponding phasor is
\begin{equation}
    \label{eq:2-sto}
    \Psi_\text{STO}[\ell] = \mathrm{e}^{-\mathrm{j} 2\pi k \nu[\ell]},
\end{equation}
with $\Psi \in \mathbb{C}$.
The quantity $\nu[\ell] \in \mathbb{R}$ denotes the time-varying sampling frequency difference between transmitter and receiver normalized by $\Delta f$

Consequently, we model the joint observation of~\eqref{eq:2-signal-model},~\eqref{eq:2-cfo}, and~\eqref{eq:2-sto} as
\begin{equation}
    \label{eq:2/joint-model}
    \widetilde{\mathbf{H}}_{k\ell} = \mathbf{H}_{k\ell} \cdot \mathrm{e}^{\mathrm{j} 2\pi \ell \mu[\ell]} \cdot \mathrm{e}^{-\mathrm{j} 2\pi k \nu[\ell]},
\end{equation}
where $\widetilde{\mathbf{H}}_{k\ell} \in \mathbb{C}$ represents the observed channel frequency response $\widetilde{\mathbf{H}} \in \mathbb{C}^{K \times L}$ at the $k$-th subcarrier and $\ell$-th symbol.

%Due to a mismatch between the underlying signal model given by \eqref{eq:2-signal-model} and the measured frequency response at the time indices where this change of phase slope occurs, model-based \gls{hrpe} methods are not capable of estimating the multipath components accurately.
%One could build this upon the fact that model-based estimators exhibit significantly reduced accuracy when there is a model mismatch.
%Consequently, we can utilize this mismatch as a metric to evaluate the compensation performance.
\section{LoS-based Synchronization}
\label{3-algorithm}

We employ a technique referred to as geometry-based synchronization for the correction of the sounding data under time-varying offsets and strong multipath propagation.
Our drift compensation algorithm comprises a \gls{hrpe} in delay domain, Kalman filtering for \gls{los} identification, and the comparison of the \gls{los} parameters to the geometry-based ground truth.
Although the error modeling is developed with consideration for \gls{gpsdo}-induced drifts, the presented correction algorithm demonstrates applicability to channel sounding data of differing synchronization and even unsynchronized measurements.

\subsection{LoS Estimation}
\label{3-algorithm/los}

We assume that all sounding data symbols contain a \gls{los} path, which can be influenced by fading.
This assumption is justified by the unobstructed outdoor measurement scenario and the omnidirectional design of the antenna beam patterns~\cite{2025_icas4m_uav_measurement_campaign_Beuster}.
To determine the \gls{los}, we first perform a \gls{hrpe} of the propagation delay for each symbol within \eqref{eq:2/joint-model} utilizing the RIMAX algorithm~\cite{2005_diss_rimax_legacy_Richter}.
Due to the model order estimation of RIMAX, this step yields $\hat{P}[\ell] \in \mathbb{N}$ multipath components, each comprising a propagation delay $\tau_p \in \mathbb{R}$ and a complex path weight~$\gamma_p \in \mathbb{C}$.
As previously stated in \Cref{2-basics/observed-effects}, the time-varying nature of \gls{cfo} and \gls{sfo} necessitate such a symbol-wise delay estimation.

\subsubsection{Simple Heuristics}
\label{3-algorithm/los/simple}

\begin{figure}[b]
    \centering
    \includegraphics[width=0.8\columnwidth]{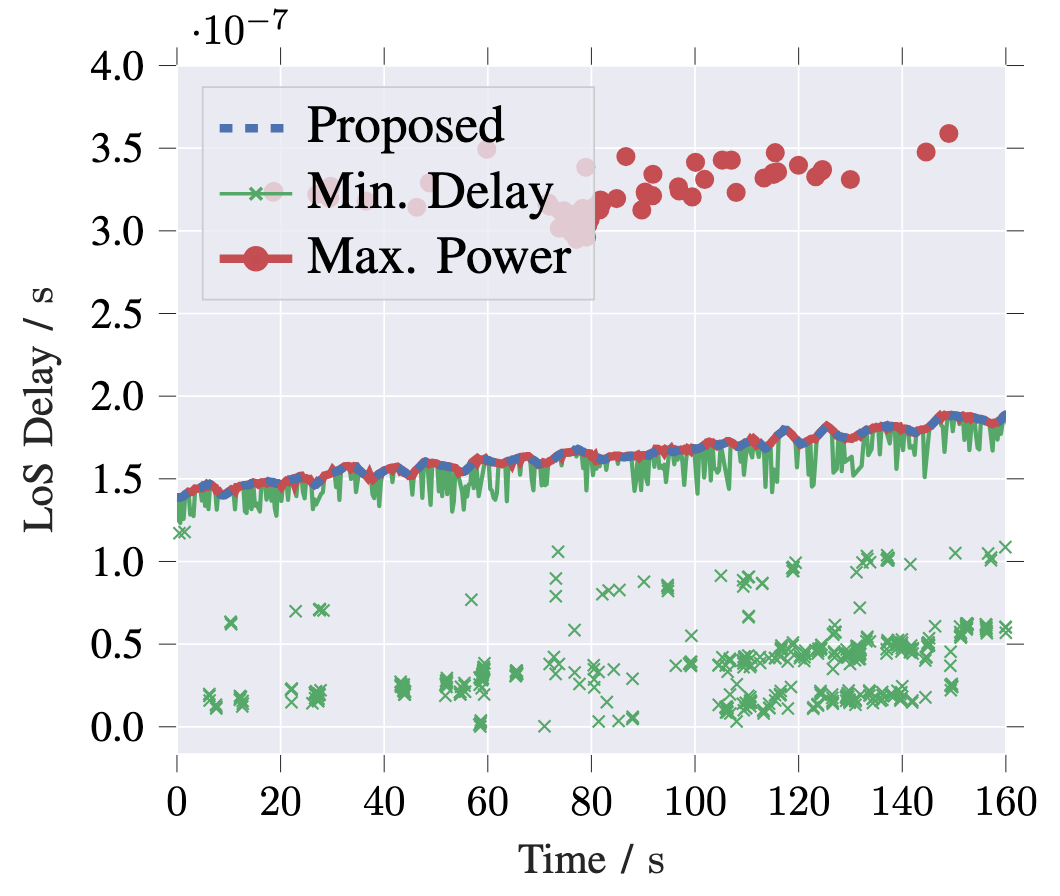}
    \caption{\textit{Comparison of Different LoS Estimations---While the proposed LoS tracking (blue) achieves a smooth progression, traditional heuristics---minimum delay (green) or maximum power (red)---fail due to pronounced multipath propagation and spurious detections.}}
    \vspace{-1em}
    \label{fig:3-los-heuristic}
\end{figure}

Simple strategies for \gls{los} estimation are the selection based on the multipath component having either the lowest delay
\begin{equation}
    \label{eq:3-mindel}
    \tilde{p}[\ell] =  \argmin_{p \in \{1, \ldots, \hat{P}[\ell]\}} \tau_p[\ell],
\end{equation}
or the highest power
\begin{equation}
    \label{eq:3-maxpow}
    \tilde{p}[\ell] =  \argmax_{p \in \{1, \ldots, \hat{P}[\ell]\}} |\gamma_p[\ell]|^2.
\end{equation}
However, as \Cref{fig:3-los-heuristic} shows, these simple heuristics repeatedly fail in the presence of rich multipath propagation due to spurious detections of the \gls{hrpe} algorithm and additional strong paths in the channel, like ground reflections.

\subsubsection{Kalman Filter}
\label{3-algorithm/los/kalman}

As the \gls{los} path within the \gls{hrpe} results tends to be unstable due to estimation variance, path splitting, temporary shadowing, and antenna beam pattern influences, we adopt a Kalman filter to provide a stable \gls{los} estimate~\cite{1960_kalman_filter_legacy}.
Both the transmitter and the receiver can undergo continuous motion.
Consequently, the delay of the \gls{los} evolves smoothly over symbols, which allows us to employ a standard constant acceleration approach with
\begin{equation}
    \label{eq:3-kalman-state-vector}
    \mathbf{x} = \left[\tau, \dot{\tau}, \ddot{\tau}\right]^\mathsf{T}
\end{equation}
denoting the corresponding state vector $x \in \mathbb{R}^3$ for the \gls{los}.

For the initialization of the Kalman filter, we select the \gls{los} of the first symbol using~\eqref{eq:3-mindel} as $\tilde{p}[0]$.
This choice is justified by the fact that the individual \gls{cfo} and \gls{sfo} phase shifts are negligible for early symbols, as their influence accumulates over time. 
The Kalman filter, then, repeatedly performs the prediction step based on the state transition matrix
\begin{equation}
    \label{eq:3-kalman-transition-matrix}
    \mathbf{F} =
    \begin{bmatrix}
        1   &   \Delta t    &   \frac{\Delta t^2}{2}    \\
        0   &   1           &   \Delta t                \\
        0   &   0           &   1
    \end{bmatrix}
    .
\end{equation}
Leveraging all $\hat{P}[\ell]$ delays, 
\begin{equation}
    \label{eq:3-kalman-innovation}
    y_p[\ell] = \tau_p[\ell] -  \mathbf{M}\mathbf{x}[\ell]  
\end{equation}
represents the innovation of the $p$-th estimate, where
\begin{equation}
    \label{eq:3-kalman-observation-matrix}
    \mathbf{M} =
    \begin{bmatrix}
        1   &   0   &   0   \\
    \end{bmatrix}
\end{equation}
denotes the measurement matrix of the filter and the vector~$x[\ell]$ is the Kalman prediction at the $\ell$-th symbol.
Out of these $\hat{P}[\ell]$ predictions, we select the one yielding the lowest Mahalanobis distance
\begin{equation}
    \label{eq:3-mahalanobis-distance}
    \tilde{p}[\ell] = \arg \min_{p \in \{1, \ldots, \hat{P}[\ell]\}} \sqrt{\frac{y_p[\ell]}{S_p[\ell]}}.
\end{equation}
Here, $S_p[\ell] \in \mathbb{R}$ denotes the variance of the $p$-th innovation, which is extracted from the error covariance matrix of the Kalman filter~\cite{1960_kalman_filter_legacy}.

After performing the Kalman update step using the \mbox{$\tilde{p}$-th} prediction, the filter outputs the delay estimate for the \gls{los} at the $\ell$-th symbol, denoted by $\hat{\tau}[\ell] \in \mathbb{R}$.
To calculate the corresponding \gls{los} path weight, we correlate the Kalman-based \gls{los} response to the measurement vector at the~\mbox{$\ell$-th} symbol, yielding
\begin{equation}
    \label{eq:3-kalman-beamform}
    \hat{\gamma}[\ell] = \sum_{k=0}^{K-1} \left(\mathrm{e}^{-\mathrm{j} 2\pi k \hat{\tau}[\ell]}\right)^* \cdot \widetilde{\mathbf{H}}_{k\ell},
\end{equation}
where $(\cdot)^*$ denotes the complex conjugate.
As \Cref{fig:3-los-heuristic} illustrates, the proposed procedure achieves a smooth \gls{los} delay progression in the multipath channel sounding measurements.

\subsection{Drift Compensation}
\label{3-algorithm/correction}

To correct for \gls{cfo} and \gls{sto}, the geometry-based algorithm utilizes the previously estimated phase and delay.
In essence, we calculate the delay and phase difference of the previously estimated \gls{los} $\hat{\tau}[\ell]$ and $\arg\hat{\gamma}[\ell]$ to its \gls{gnss}-\gls{rtk}-derived ground truth $\tilde{\tau}[\ell]$ and $\arg\tilde{\gamma}[\ell]$, respectively.
Applying these differences to the uncompensated data yields
\begin{equation}
    \label{eq:3-joint-correction}
    \hat{\mathbf{H}}_{k\ell} = \widetilde{\mathbf{H}}_{k\ell} \cdot e^{-j 2\pi \left(\arg\hat\gamma[\ell] - \arg\tilde\gamma[\ell]\right) } \cdot e^{j 2\pi k \left(\hat\tau[\ell] - \tilde\tau[\ell]\right)},
\end{equation}
the drift-compensated frequency response $\mathbf{\hat{H}} \in \mathbb{C}^{K \times L}$ of the $k$-th subcarrier and $\ell$-th symbol.

\section{Results and Analysis}
\label{4-results}

We explicitly tailor the introduced algorithm to compensate synchronization mismatches of the sounding data presented in~\cite{2025_icas4m_uav_measurement_campaign_Beuster}.
The compensated sounding measurements utilizing the presented post-processing are publicly available at~\cite{2025_oryx_dataset_Miranda}.
For evaluation purposes, we perform 2-D \gls{hrpe} of the delay-Doppler parameters on the uncompensated and compensated data with coherent processing intervals of \qty{0.18}{\second} which is the equivalent of \num{562} symbols.
This step yields estimates denoted by $\boldsymbol{\theta}[s] \in \mathbb{C}^{\hat{P}[s]\times 3}$ for the $s$-th processing interval each comprising $\hat{P}[s]$ propagation delays $\hat{\tau}_p[s] \in \mathbb{R}$, Doppler-frequencies $\hat{\alpha}_p[s] \in \mathbb{R}$, and path weights $\hat{\gamma}_p[s] \in \mathbb{C}$.

\subsection{Evaluation Metrics}
\label{4-results/metric}

The analysis of the drift compensation based on different \gls{los} estimates requires the definition of appropriate metrics.
We pursue a combination of two strategies for the evaluation.

\subsubsection{Relative Residual Power}
\label{4-results/metric/relres}

The first metric is the so-called relative residual power defined by
\begin{equation}
    \label{eq:5-relrespower}
    \epsilon(\widetilde{\mathbf{H}}, \mathbf{H}, \boldsymbol{\theta}) = \frac{\lVert \widetilde{\mathbf{H}} - \mathbf{H}(\mathbf{\boldsymbol{\theta}})\rVert_2^2}{\lVert \widetilde{\mathbf{H}} \rVert _2^2},
\end{equation}
where $\mathbf{H(\boldsymbol{\theta})}$ denotes the noiseless version of \eqref{eq:2-signal-model} utilizing the previously obtained \gls{hrpe} estimates~$\boldsymbol{\theta}$ of the $s$-th processing interval~\cite{2023_hrpe_wideband_sounding_Semper}.
This quantity measures the power of~\eqref{eq:2/joint-model} not described by~\eqref{eq:2-signal-model} and does not require any ground truth knowledge about the measurement data.

As the parametric signal model~\eqref{eq:2-signal-model} is closely related to physics, we expect a smooth phase within the measurements.
The observed non-smooth phase progressions within a coherent processing interval are detrimental to the performance of model-based estimation routines, as the resulting phase incoherencies are not comprised in the underlying parametric signal model~\eqref{eq:2-signal-model}.
This circumstance makes the relative residual power a proxy to quantify the degree of incoherency reduction by the compensation algorithms.

%Since \cref{eq:2-signal-model} comprises linear phase variations, the observed drift in delay and Doppler within \cref{fig:2-sync-implications-1} can be explained by the model and therefore should not affect this metric.
%In contrast, fast phase changes during a coherent processing interval that cause the smearing effects in \cref{fig:2-sync-implications-2} are not part of the signal model.
%As a result, such coherent processing intervals with smearing yield increased relative residual power due to a mismatch between the signal model and the actual data.

%The relative residual power does not require any ground truth knowledge about the measurement and, instead, is purely data-driven.
%Since a lower residual power implies a better synchronization of the data, this metric enables us to compare the different \gls{los} estimation approaches introduced in \cref{3-algorithm/los}

\subsubsection{RMSE}
\label{4-result/metric/rmse}

In contrast to the relative residual power, the \gls{rmse} of the estimates of delay and Doppler parameters is relevant for applications like sensing.
Using the \gls{gnss}-\gls{rtk}-derived delay-Doppler parameters of passive objects---also called targets---we compare how the post-processing affects these parameters.
The resulting \gls{rmse} of~$T$ known targets is defined as the corresponding geodesic distance on $S^1$ with
\begin{equation}
    \mathrm{RMSE}_\xi = \sqrt{\frac{1}{T} \sum_{t=1}^T \min \left(|\hat{\xi}_t - \tilde{\xi}_t|,\ 1 - |\hat{\xi}_t - \tilde{\xi}_t| \right)^2},
\end{equation}
where $\xi_t \in \mathbb{R}$ represents a placeholder for either delay $\tau_t$ or Doppler $\alpha_t$ of the $t$-th target.
The association between estimations $\hat{\xi}_t$ and ground truths $\tilde{\xi}_t$ is performed by minimizing the distance between them over all permutations of the joint distances between the corresponding delay-Doppler parameter sets.
We achieve this minimization by employing Hungarian matching~\cite{1955_hungarian_algorithm_legacy_Kuhn}.

\subsection{Compensation of the Channel Sounding Data}
\label{4-results/sounding}

\begin{figure}[h]
    \centering
    \begin{subfigure}{1\columnwidth}
    \centering
    \includegraphics[width=0.8\columnwidth]{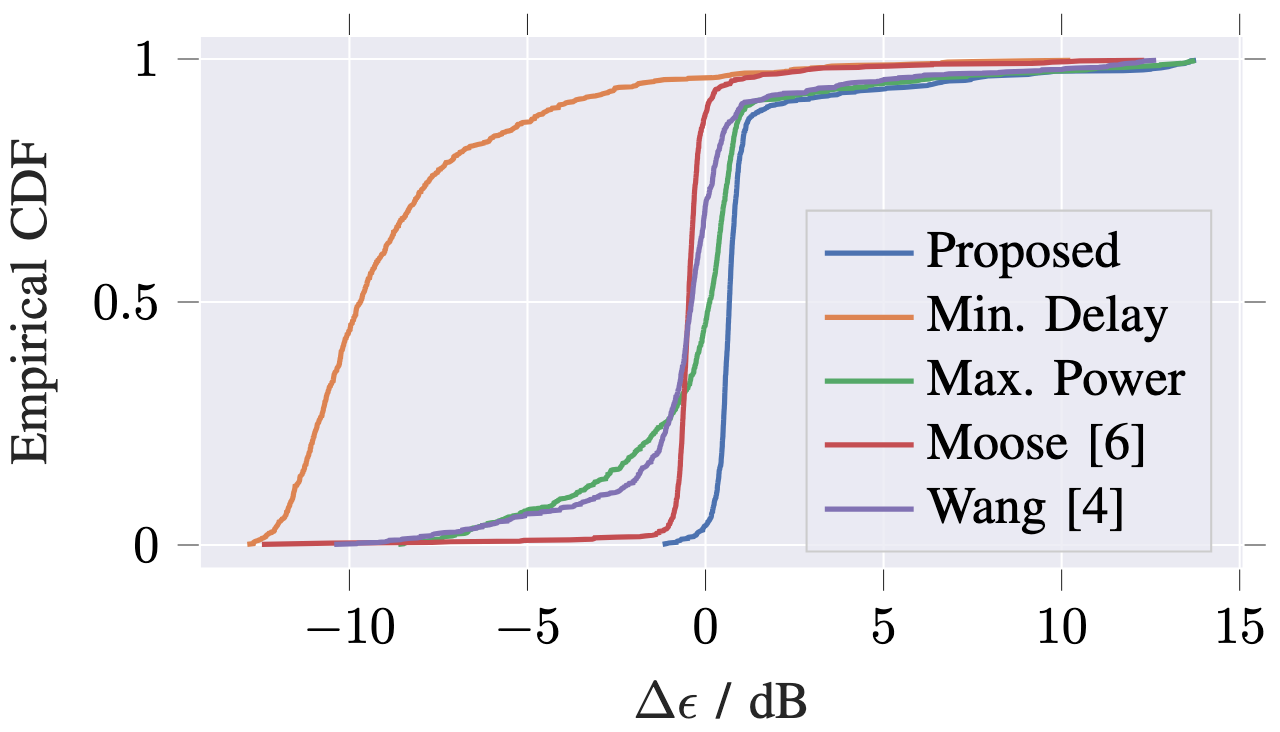}
    \caption{\textit{Single Node Relative Residual Power Reduction $\Delta \epsilon$}}
    \label{fig:4-cdf-single}
    \end{subfigure}
    %\hfill
    \begin{subfigure}{1\columnwidth}
    \centering
    \includegraphics[width=0.8\columnwidth]{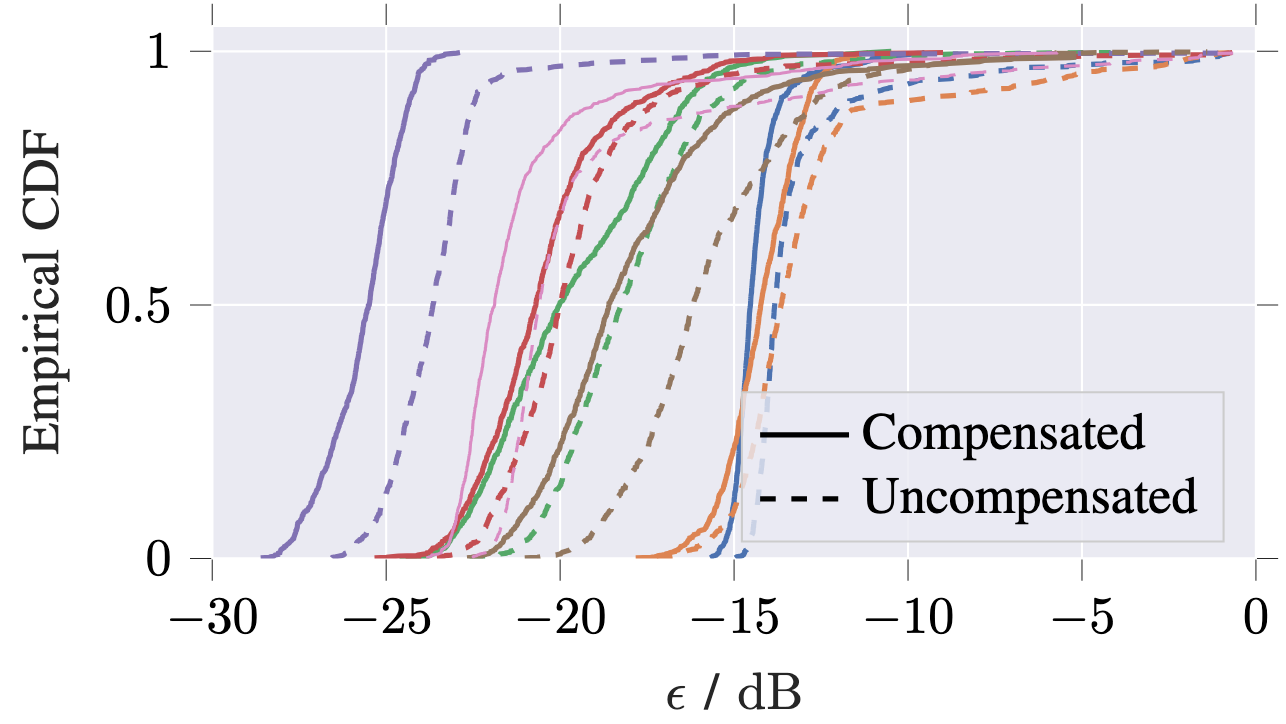}
    \caption{\textit{Multi-Node Relative Residual Powers $\epsilon$}}
    \label{fig:4-cdf-multi}
    \end{subfigure}
    \caption{\textit{Empirical CDF of the Relative Residual Power for the Multi-Node Sounding Setup---The proposed LoS tracking achieves the best reduction on the measurements (a) thus reducing model mismatch of multi-node measurements reliably (b).}}
    \vspace{-2em}
    \label{fig:4-cdf}
\end{figure}

As \Cref{fig:4-cdf-single} illustrates by the empirical \gls{cdf} for one receiver, our proposed extension of geometry-based compensation (blue) reduces the relative residual power for approximately~\qty{95}{\percent} of the processing intervals.
%The Kalman-based \gls{los} compensation reduces this metric for each coherent processing interval by removing incoherencies due to non-smooth phase progressions.
In contrast, the two \gls{los} heuristics (orange, green) increase the relative residual power.
Multipath propagation and spurious detections of the \gls{hrpe} delay estimation result in an alternation of the estimated \gls{los} between the actual path and these reflections, which leads to fluctuations in both delay and phase and adds additional incoherencies to most processing intervals.
The two implemented reference algorithms of Moose~\cite{1994_cfo-correction_Moose} and Wang~\cite{wang_system_error_2022} also increase incoherencies reflecting in overall increased relative residual powers.
These two algorithms do not explicitly estimate the \gls{los} path and, instead, determine \gls{cfo} and \gls{sfo} based on a reference symbol of the frequency response
Consequently, both approaches fail a reliable compensation due to the time-variance of these offsets and \gls{los} fading.

\Cref{fig:4-cdf-multi} illustrates the empirical \gls{cdf} of the relative residual power prior and post to compensation for each receiver of the multi-node channel sounding data.
The Kalman-based \gls{los} tracking in combination with geometry-based compensation reliably reduces the relative residual power across all nodes.
In particular, processing intervals containing the highest incoherencies experience a reduction of this metric by at least~\qty{5}{\decibel}.

\begin{figure}[h]
  \centering
  % First subfigure
  \begin{subfigure}[b]{0.46\columnwidth}
    \centering
    \includegraphics[width=1\columnwidth]{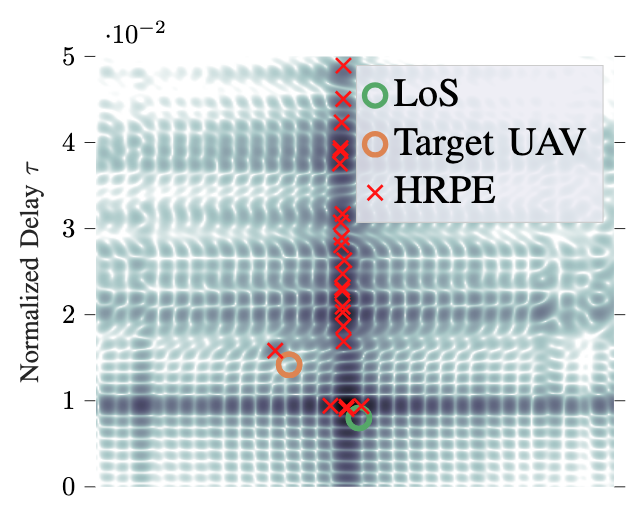}
    \caption{\textit{Drifts Uncompensated}}
    \label{fig:4-ddmap-shift}
  \end{subfigure}
  \hfill
  % Second subfigure
  \begin{subfigure}[b]{0.46\columnwidth}
    \centering
    \includegraphics[width=0.9\columnwidth]{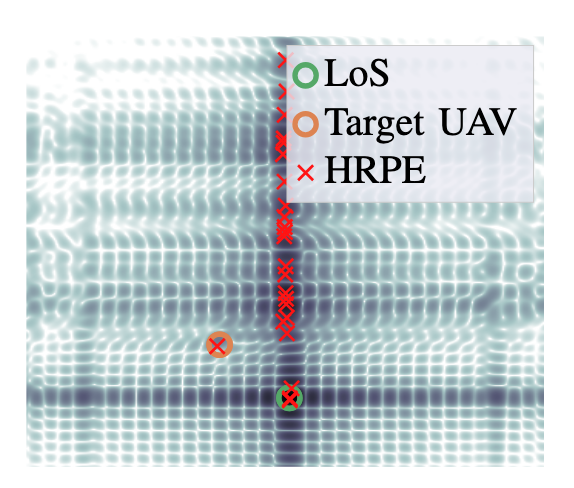}
    \caption{\textit{Drifts Compensated}}
    \label{fig:4-ddmap-shift-corrected}
  \end{subfigure}
  \begin{subfigure}[b]{0.46\columnwidth}
    \centering
    \includegraphics[width=1\columnwidth]{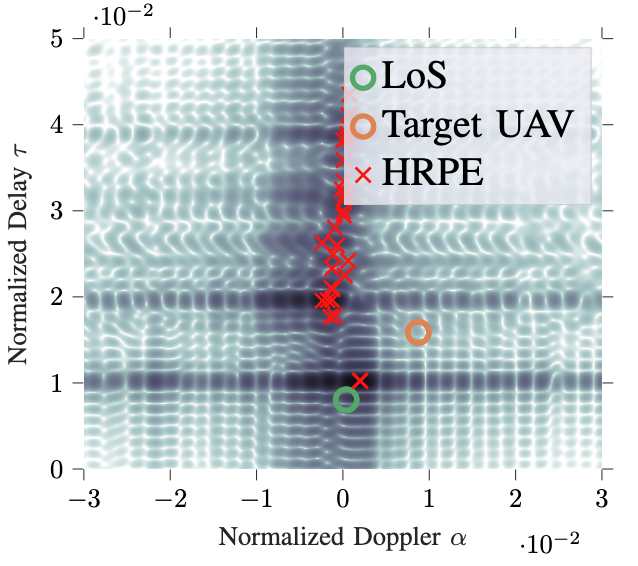}
    \caption{\textit{Incoherence Uncompensated}}
    \label{fig:4-ddmap-smear}
  \end{subfigure}
  \hfill
  \begin{subfigure}[b]{0.46\columnwidth}
    \centering
    \includegraphics[width=0.9\columnwidth]{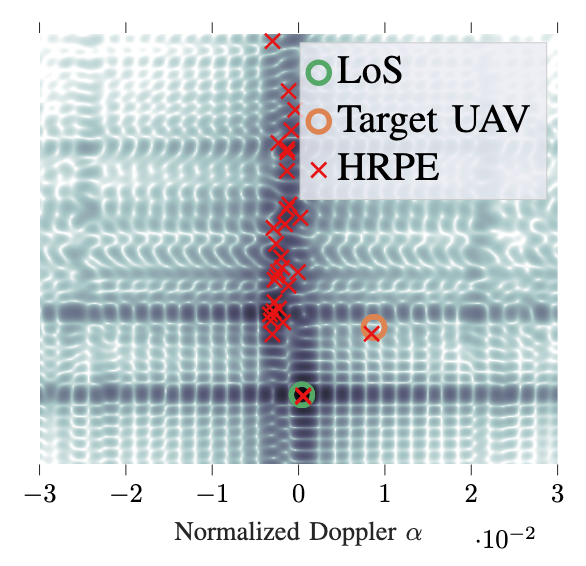}
    \caption{\textit{Incoherence Compensated}}
    \label{fig:4-ddmap-smear-corrected}
  \end{subfigure}
  \hfill
  \begin{subfigure}[b]{0.46\columnwidth}
    \centering
    \includegraphics[width=1\columnwidth]{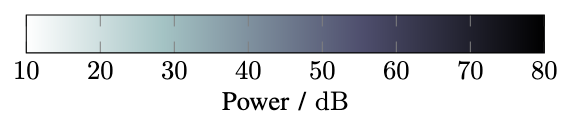}
  \end{subfigure}
  \caption{\textit{Selected Delay-Doppler Spectra of the Measurement Data Prior and Post to Compensation---The uncompensated data exhibits time-varying drifts in delay-Doppler (a) and incoherence (c). Applying the proposed LoS-based correction effectively removes both effects ((b), (d)) and makes the target detectable~(d).}}
  %The proposed compensation removes drifts~((a)\textrightarrow(b)) and incoherence~((c)\textrightarrow(d)).}
  \vspace{-1.5em}
  \label{fig:4-ddmap}
\end{figure}

\Cref{fig:4-ddmap} depicts delay-Doppler spectra of two cherry-picked coherent processing intervals prior and post to compensation.
Applying the proposed approach removes delay-Doppler drifts~(\Cref{fig:4-ddmap-shift} $\text{\textrightarrow}$ \Cref{fig:4-ddmap-shift-corrected}) and reduces incoherencies~(\Cref{fig:4-ddmap-smear} $\text{\textrightarrow}$ \Cref{fig:4-ddmap-smear-corrected}).
As \Cref{fig:4-ddmap-smear-corrected} demonstrates, the reduction of the latter enables the \gls{hrpe} algorithm to detect a passive \gls{uav} target that was ``hidden'' within the uncompensated data. 

In addition, \Cref{tab:4-rmse} lists the delay-Doppler \glspl{rmse} of this target.
While the proposed algorithm decreases these errors by approximately~\qty{60}{\percent} compared to the uncompensated channel sounding data, the other algorithms demonstrate no reliable reduction of the \glspl{rmse}.
Consequently, the Kalman-based compensation increases the consistency between the delay-Doppler estimates and their geometrically-derived ground truth of our sounding data, improving passive target localization, for example.

Overall, the proposed compensation enhances the quality and usability of our \gls{gpsdo}-based sounding data~\cite{2025_oryx_dataset_Miranda}.
Moreover, the algorithm is suitable for post-processing synchronization of arbitrary channel sounding data suffering time-varying drifts in strong multipath environments.
The only two requirements are accurate position information of the transceiver nodes and an existing but not necessarily dominant \gls{los} path.

\begin{table}[tb]
\scriptsize
\setlength{\tabcolsep}{2pt} % Default is 6pt
\sisetup{propagate-math-font = true, reset-math-version = false}
\centering
\caption{\textit{Target RMSEs of Synthetic and Sounding Data}}
\begin{tabular}{c c c c c c c}
\toprule
& \multicolumn{6}{c}{Correction Algorithm}\\
RMSE & Uncompensated & \textbf{Proposed} & Min. Delay & Max. Power & Moose \cite{1994_cfo-correction_Moose} & Wang \cite{wang_system_error_2022}  \\
\midrule
%\multicolumn{7}{c}{Synthetic Data}\\
%\midrule
%Delay & \qty{1.02}{\nano\second} & \boldmath{\qty{0.06}{\nano\second}} & \qty{0.31}{\nano\second} & \qty{0.11}{\nano\second} & \qty{1.52}{\nano\second} & \qty{2.06}{\nano\second} \\
%Doppler & \qty{5.63}{\hertz} & \boldmath{\qty{1.33}{\hertz}} & \qty{1.79}{\hertz} & \qty{1.36}{\hertz} & \qty{2.94}{\hertz} & \qty{4.74}{\hertz} \\
%\midrule
%\multicolumn{7}{c}{Sounding Data}\\
%\midrule
Delay & \qty{14.76}{\nano\second} & \boldmath{\qty{6.06}{\nano\second}} & \qty{28.16}{\nano\second} & \qty{36.74}{\nano\second}  & \qty{16.31}{\nano\second} & \qty{20.25}{\nano\second}\\
Doppler & \qty{4.61}{\hertz} & \boldmath{\qty{1.68}{\hertz}} & \qty{2.06}{\hertz} & \qty{3.63}{\hertz} & \qty{3.82}{\hertz} & \qty{4.33}{\hertz} \\
\bottomrule
\end{tabular}
\label{tab:4-rmse}
\end{table}
%\glsresetall
\section{Conclusion}
\label{5-conclusion}

Within our work, we extend geometry-based drift compensation algorithms for post-processing of distributed multi-node channel sounding data by employing a symbol-wise \gls{hrpe} algorithm and a Kalman filter for explicit \gls{los} determination.
We demonstrate that this extension facilitates the estimation and correction of time-varying \gls{cfo} and \gls{sfo} under rich multipath propagation from the measurements. %while preserving phase changes due to relative transmitter-receiver movement.
In addition, we show that the relative residual power is a suitable metric to quantify the compensation performance of different drift correction algorithms.
Evaluating the synchronization quality on the sensing accuracy of a passive target, we are capable of reducing the delay-Doppler \glspl{rmse} by \qty{60}{\percent} and relative residual powers by at least \qty{5}{\decibel}.

Although the developed algorithm itself is not suitable for real-time implementation, the concept of combining an explicit \gls{los} estimation step with a Kalman filter has the potential to facilitate the distributed synchronization of operational communication and sensing systems.
As these systems already employ synchronization utilizing dedicated signals, future research must assess whether the remaining offsets are still detrimental to sensing.
In this scenario, the geometry-based synchronization bears the potential to further improve estimation accuracy in distributed systems.

%The presented algorithm is designed to perform drift correction on recorded channel sounding data.
%Our approach yields sounding data with coarse coherence, a continuously differentiable phase progression over a large time-scale, comparable to that observed in practical communication systems.
%Moreover, future operational \gls{icas} systems will require such smooth phase progression over much shorter time-scales, known as coherent processing intervals.
%The coarsely compensated data obtained with our \gls{los} tracking approach, therefore, offers a promising foundation for research on operational phase correction algorithms for \gls{icas}.

%The presented algorithm is tailored to perform drift correction on recorded channel sounding data.
%This compensation procedure is paramount for ensuring a continuously differentiable phase progression over a large time-scale, comparable to the coherence of actual communication systems.
%Consequently, our work highlights that a continuoperational \gls{icas} systems require a similar synchronization accuracy of their nodes.
%
%Here, the proposed \gls{los} tracking constitutes a promising approach for distributed synchronization of operational systems.

%However, future \gls{icas} systems are likely to require precise real-time synchronization for high-accuracy sensing.
%Improving the presented \gls{los} estimation in terms of real-time capability is a crucial next step towards accurate and reliable operational \gls{icas} systems.

\section*{Acknowledgment}
This work is sponsored by the BMFTR project 6G-ICAS4Mobility with Project No. 16KISK241.
We would also like to thank Michael Döbereiner and Reza Faramarzahangari for lending their technical expertise during this work.
\printbibliography
% carstens bst file
%\bibliographystyle{IEEEtran}
%\bibliography{literature}

\end{document}